# Preparing data for pathological artificial intelligence with clinical-grade performance

Yuanqing Yang[1], Kai Sun[1], Yanhua Gao[2], Kuangsong Wang[3,4], Gang Yu[1, *]

1 Department of Biomedical Engineering, School of Basic Medical Sciences, Central South University, Changsha, China

2 Department of Ultrasound, Shaanxi Provincial People's Hospital, Xi'an, China

3 Department of Pathology, School of Basic Medical Sciences, Central South University, Changsha, China

4 Department of Pathology, Xiangya Hospital, Central South University, Changsha, China

* Correspondence author

Gang Yu, Email：yugang@mail.csu.edu.cn

## Abstracts

**[Purpose]** The pathology is decisive for disease diagnosis, but relies heavily on the experienced pathologists. Recently, pathological artificial intelligence (PAI) is thought to improve diagnostic accuracy and efficiency. However, the high performance of PAI based on deep learning in the laboratory generally cannot be reproduced in the clinic.

**[Methods]** Because the data preparation is important for PAI, the paper has reviewed PAI-related studies in the PubMed database published from January 2017 to February 2022, and 118 studies were included. The in-depth analysis of methods for preparing data is performed, including obtaining slides of pathological tissue, cleaning, screening, and then digitizing. Expert review, image annotation, dataset division for model training and validation are also discussed. We further discuss the reasons why the high performance of PAI is not reproducible in the clinical practices and show some effective ways to improve clinical performances of PAI.

**[Results]** The robustness of PAI depend on randomized collection of representative disease slides, including rigorous quality control and screening, correction of digital

discrepancies, reasonable annotation, and the amount of data. The digital pathology is fundamental of clinical-grade PAI, and the techniques of data standardization and weakly supervised learning methods based on whole slide image (WSI) are effective ways to overcome obstacles of performance reproduction.

**[Conclusion]** The representative data, the amount of labeling and consistency from multi-centers is the key to performance reproduction. The digital pathology for clinical diagnosis, data standardization and technique of WSI-based weakly supervised learning hopefully build clinical-grade PAI.

# 1. Introduction

Pathological diagnosis is the gold standard for disease diagnosis[1], which is diagnosed by the tissues and cells in the samples on glass slides using the microscope. Glass slides are inconvenient to storage, sharing and remote consultation. With development of digitalization technology, the pathological scanner converts the slide to high-resolution whole slide image (WSI)[2], and pathological diagnosis based on digital image is becoming popular[3,4], which is also called digital pathology (DP).

Recently, the combination of DP and artificial intelligence (AI) has given birth to the new computational pathology (CPATH) or pathological AI (PAI), improving the efficiency and accuracy of diagnosis and alleviating the shortage of pathologists. For example, the current shortage of pathologists in China is up to 100,000, in addition, the number of pathologists in the United States also decreased by 17.53% from 2007 to 2017[5]. In the past few years, a large number of PAI systems have emerged for the classification, grading, outcome prediction, prognosis determination [6,7] and cancer diagnosis such as gastric cancer[8,9], prostate cancer [10-13], bowel cancer[14], breast cancer[15-19], cervical cancer[20,21].

The PAI is generally based on deep learning, train and test on the dataset with hundreds to tens of thousands of WSIs. Although the PAI has proved to be effective and robust, showing extremely high performance, even comparable to pathologists, their high performance is generally difficult to reproduce in the clinic. The data preparation for training and testing greatly affects the accuracy and generalization ability of an AI based on deep learning, which is also the key to overcome the obstacles for PAI applied in clinical practice[22].

Based on the extreme importance of dataset to develop effective PAI, we reviewed the PAI studies published before, and then discussed data preparation methods for accurate and robust AI to improve the PAI study in the future. The main motivation is to find a solution to the obstacles to the clinical application of PAI, that is, how to prepare data for developing clinical-grade PAI.

A search was conducted using the PubMed database with the time frame set from January 2017 to February 2022, using the keywords such as pathology, machine learning, digital pathology, pathological diagnosis, and deep learning. A total of 829 papers were retrieved. Figure 1 shows the number of PAI studies using deep learning methods in the last 5 years. Among them, there are the highest number of 450 studies published in 2021 shown in Figure 1.

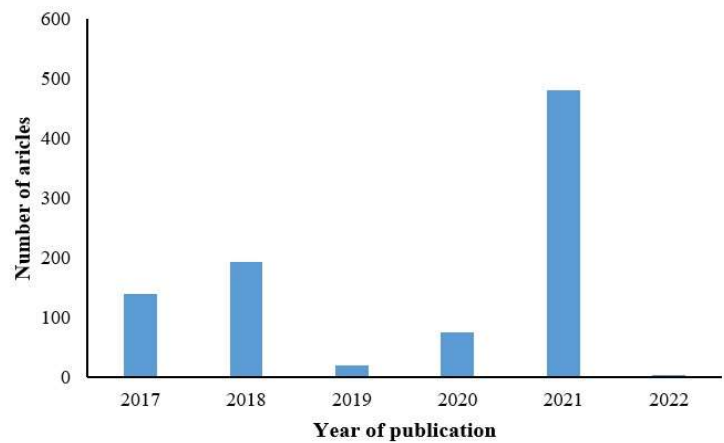

Figure 1. Publications of PAI by years

We had read abstracts of the papers, reviewed research types and topics, and screened out 220 papers related to pathological images. After reading the full text, 118 papers met the goals of data preparation analysis with the following three rules for exclusion.

1. The data used was not pathological image.

2. The work hadn't described the data preparation process in details or used only public datasets.

3. The study was only related to AI or only related to pathology.

We provided an in-depth analysis of the 118 works on the pipeline of data preparation for building effective PAI systems, as shown in Figure 2. The first step is to collect tissue samples according to the PAI objectives. The dust is removed from slides and tissue color and integrity are then checked. These slides are digitized by a pathology scanner to obtain WSIs. The quality controlling of WSIs is essential for excluding the artifacts in images, and various methods of annotation are used to relate the data and medical facts. After model training, general validation should be performed to evaluate accuracy and generalization of PAI.

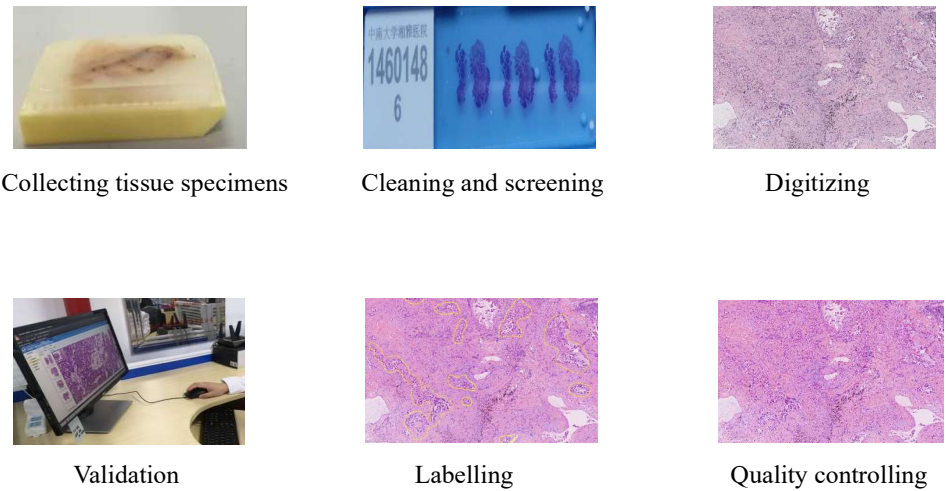

| Collecting tissue specimens | Cleaning and screening | Digitizing |
| Validation | Labelling | Quality controlling |

Figure 2 Data preparation pipeline for robust PAI.

## 2. Data collection

**2.1 Collection for pathological slides**

PAI requires a large number of high-quality WSIs from clinical diagnoses[23], but the tissue specimens exist in the form of slides. Researchers have to determine the keyword for search, and the technicians use keyword to gain access to patient identity from Pathology Information System (PIS), and obtain the required slides.

Deep learning performance is closely relied on the amount of training data. For example, as the dataset of colorectal cancer was increased from 420[24] to 13,111 slides [25], the accuracy of PAI significantly increased, and even exceeded that of pathologists, showing good multicenter generalization ability[26-28]. The 44,732 slides from 15,178 cases were collected and area under curve (AUC) even achieved as high as 0.98 on independent medical centers[23]. In addition to the number of slides, disease representation included in the slides is also important, such as all disease subtypes and grades etc. so as to avoid the data bias[28].

The balance of the number of slides from various classes is another issue that should be paid special attention[20]. Some diseases are relatively rare, thus resulting in a low number of slides. The imbalance dataset of renal cell carcinoma resulted in 10% lower recall for the disease type with small sample size than the remaining subtypes[29]. If the age, gender, and post-surgery outcomes may affect the results, the number of slides for each factor should be close. One way to keep data balanced is to randomly select a portion from slides with a class with large sample size. For example, in order to equalize the number of survivors and deaths of colorectal cancer patients within 5 years after surgery, 182 survivor data were randomly removed [24]. It is also feasible to increase slide number of class with small sample size from other centers.

In order to avoid the influence of uncontrollable factors, the collection of slides should be as randomized as possible[14]. For example, reagent types and preparation method of slides may be the same in a certain period of time, different in time periods. The span of time when the slide was made should be as long as possible to include variations in the production process. The 250 specimens from January 2009 to December 2017 were collected, ensuring AUC of hepatocellular carcinoma prognostic model reach 0.95 [30]. As the span of disease-free survival and age were as long as 4-86 months and 25-75 years respectively, the prognostic model of oral squamous cell carcinoma achieved high accuracy up to 96.31% [31]. However, extensive assessments of potentially influential factors such as ethnicity, brand of tableting drug, etc. are lacking.

It is a good idea to collect multi-center slides [8,22,32-35], which are different in production protocols and drugs. PAI trained on one center dataset may suffer from a significant degradation on other centers[36], while multi-center training set may alleviate it. A model using the dataset from multiple centers has an average Dice coefficient of 5.6% [22]. The slides from three hospitals (different production protocols, four pathological scanners) improved the AUC of the model from 0.808 to 0.983 [20]. In addition, slides from multiple centers allow for a more comprehensive generalization assessment, discussed in Section 6.

Public datasets such as the TCGA (The Cancer Genome Atlas) can be used as a supplement. Many PAIs are trained and tested by public datasets [16,25,26,30,37-40], and there are also some challenges that provide pathological images, such as the Grand-challenge (https://camelyon.grand-challenge.org) [41], MITOS-ATYPIA[42] etc. listed in Table 1.

Table 1. Public datasets for pathology AI

| Dataset | Number | Disease type | Reference |
| --- | --- | --- | --- |
| PatchCamelyon | 400 WSI | lymph nodes | [43] |
| GlaS | 165WSI | Stage T3 or T4 colorectal adenocarcinoma | [44] |
| LUAD-HistoSeg | - | lung cancer | [45] |
| Camelyon16 Dataset | 400 WSI | lymph nodes | [41] |
| Colorectal cancer | 10 WSI+5000 patches | bowel cancer | [46] |
| PAIP | 100WSI | liver | [47] |

However, the images in public datasets are usually derived from a small number of slides. For example, the PAIP dataset contains only one hundred WSIs. Second, the staining process of the same WSI is similar, so the patches cut from the same WSI are also similar. Third, the public datasets may be only applicable to specific diseases, for example, the GlaS contains only images of T3 or T4 colorectal adenocarcinoma.

**2.2 Ethics Statement**

The ethical approval by the local ethics committee is required for PAI study. Although most studies are retrospective, and informed consent from patients is not required [9,24,48], however, for some prospective studies [11] such as disease outcomes [31,49], private information (name, date of birth and so on) should be anonymized [16]. The National Management Measures for Health Care Big Data Standards, Security and Services (Trial) [50], Health Insurance Portability and Accountability Act (HIPAA) [51] and other related laws should be complied strictly in data collection, storage, usage and disclosure.

**2.3 Screening and review of slides**

As many slides have been stored for a long time, the selected slides need to be cleaned to remove contamination. It is difficult to maintain a consistently high quality of production for a large number of tissue specimens. After cleaning, the quality of the slides must be carefully checked under microscope, including the integrity of the tissue specimen, tissue folding, air bubbles, staining quality and fade or not [33].

When the specimens were folded and wrinkled, the mean absolute error between the immunohistochemical score calculated by the model and the results predicted by the pathologist was 2.24 higher[52] . Therefore, unqualified slides should be excluded before the next. In addition, since the misdiagnose always happened, it should be ensured that the diagnostic results of the collected slides are correct[32]. So, the slides should be reviewed again now or after digitization.

## 3. Digitalization and quality control

**3.1 Digitization**

High-resolution WSIs are obtained from slides by a fully automatic pathological scanner[53], providing a wealth of information about the morphological and functional characteristics of biological systems [54,55]. There are various file formats of WSIs, for example, KFB file from Ningbo Jiangfeng Bio-Information Technology Co[32] , Leica 's SVS file[30] and TIFF file[23] , causing some trouble for data sharing [58]. The files should be converted to common image formats such as JPEG by the software library of manufacturer.

Recently, more and more manufacturers produce pathological scanners. However, there are always differences in the obtained WSIs [56], affecting PAI performance. For example, a Philips Ultra-Fast scanner was used to rescan prostate cancer slides scanned by an Olympus VS120-S5, resulting in 5% increase of AUC[57].

The magnification of WSI has a significant impact on model performance [106]. While scanning slides with lower magnification such as 5X (times), the cell morphology is not easily observed, but macroscopic tissue structure can be seen. Enough details can be seen from WSI of higher magnification, but more redundant pixels are included, leading to large amount of computation to AI model (Figure 3). Usually, 20X [25,59,60] and 40X [20,30,37,61] WSIs are used for most PAIs.

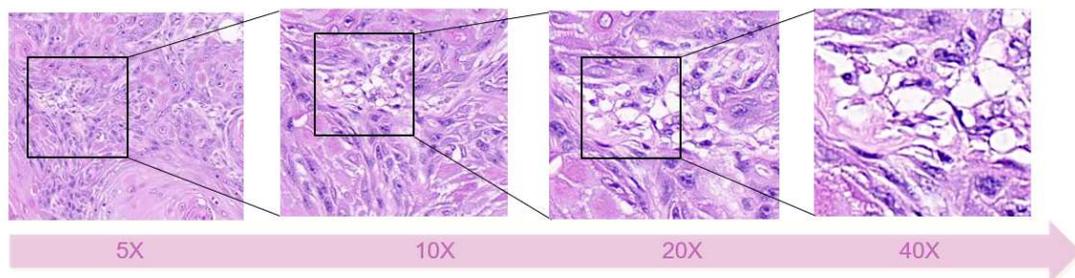

Figure 3. WSIs of different magnifications (5X, 10X, 20X to 40X from left to right.)

To overcome the differences of scanners during digitization, recent studies have shown the scanning using a variety of different scanners[7,16,20] can reduce influences from different sharpness, resolution, and imaging differences on PAI. The magnification, color fidelity and imaging quality of scanners should be carefully evaluated. The Digital Pathology Commission of the Federal Association of German Pathologists has developed a guideline on the digitization of pathology[62], which describes the minimum technical requirements for scanner systems that can be used in digitalization. Due to lack of general standards and specifications, more effectively calibrated evaluation system is urgently needed to ensure the validity of the scanners.

**3.2 Post-processing after digitalization**

The quality of some WSIs may not be ideal for developing PAIs. The tissue folding, poor staining and other problems in slides may be introduced into the images [8]. The digitalization may be failure, for example defocusing[53] as shown in Figure 5. The main distortion in WSI lies in color change, partial out-of-focus and noises. Studies have shown that the accuracy of PAI was reduced by 6%-22.8% on the images without image normalization[63]. Therefore, the post-processing after digitalization is necessary to improve PAI performance.

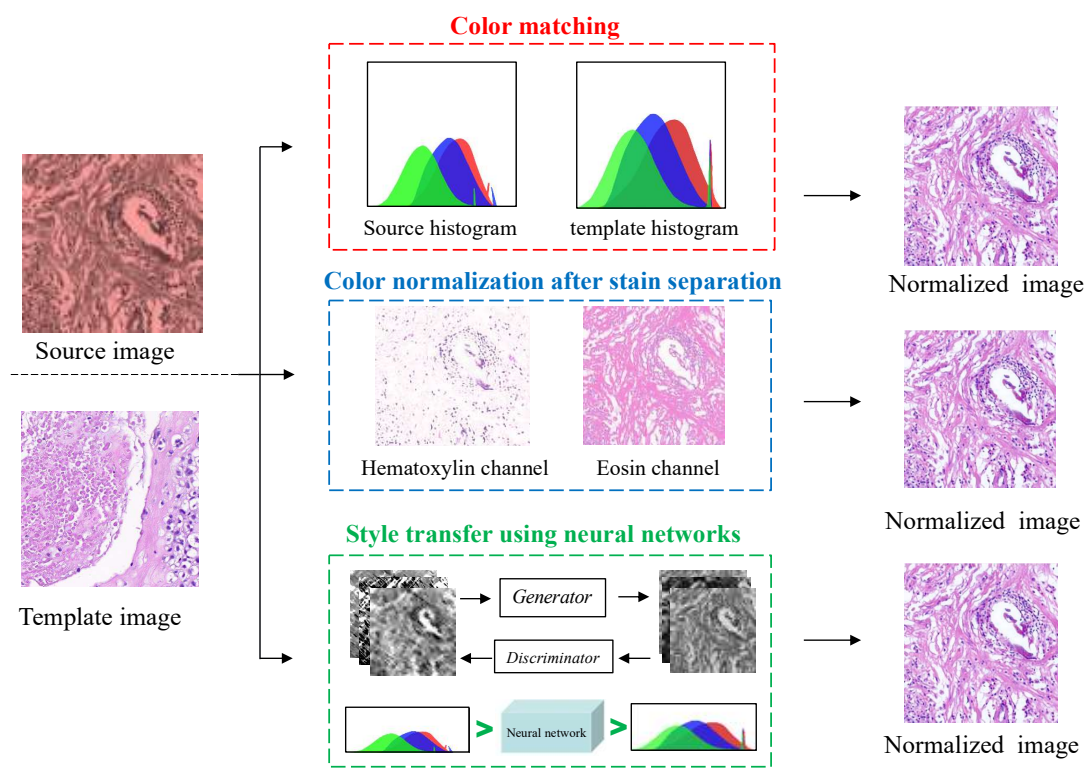

Figure 4. Three methods of color normalization. From top to bottom: The red box is the histogram-based color matching. Color normalization after stain separation is shown in the blue box, where the contribution of individual stain is separated, and the input image is matched to the template image. The style transfer method is shown in green box, where the color style of the source image is converted to that of the template image.

### 3.2.1 Color normalization

The proposed methods of color normalization can be classified into color matching, color normalization after stain separation and neural networks for style transfer, shown in Figure 4. The color matching aligns the statistical color and intensity distributions (e.g. mean and standard deviation) between a source image and a pre-selected target image[64], where the histogram specification is the most commonly used[65]. However, this method uses contrast stretching forcing the histogram of source image to match the histogram of destination image, resulting in unnatural effects[66], which may lead to unnecessary bias to subsequent image analysis[67].

The staining separation method normalizes each color channel separately. Since the relationship between the concentration of the RGB dyes and the light intensity is non-linear, it is impossible to directly use RGB for dye separation. Therefore, the RGB channels are converted to optical density（OD）space, where dye concentration and light intensity is linearly separable[66]. The image intensity (V) is defined as the logarithm of the ratio of incident ($I_0$) to transmitted light intensity ($I$) :

$$V = log10(\frac{I_0}{I}) = W \cdot H \tag{1}$$

The OD value is the staining vector ($W$) times the staining density map ($H$). Recent studies use neural networks to automatically estimate the correct $W$ and then perform a deconvolution operation for image construction [68-71].

The coloring normalization has been gradually transformed into the style transfer [72], which is learned by a generative network. The input image is migrated to the coloring style of the target image as much as possibly to restore the normal color features[63]. This method makes the input image similar to target image without the reference image. More recently, the conditional generative adversarial network (cGAN ) for color normalization reduces the randomness of manual selection and overcomes the limitation of only learning one color style in previous studies[93].

**3.2.2 Image distortion correction**

The bubbles and tissue folding in slides will lead to artifacts in WSIs[73] . The saturation changes in the folded tissue (Figure 5a) may degrade the PAI performance. The folded regions can be detected by enhanced brightness of image pixels, but some isolated pixels may be mislabeled[74] . If the connectivity of saturation and intensity was used for folding detection in low resolution WSI, AUC improved by 5% after excluding the regions with tissue folding[75].

Because the higher magnification lens in the scanner has a smaller depth of field, the WSI could be out-of-focus for uneven tissue in the slide, resulting in blurred images. Although the imaging algorithm tries to adjust focus position, local blur always occurs due to scan speed limitations [76]. Blurred image may affect the detection and classification of some diseases, where the out-of-focus normal tissue may be misjudged as the tumor [77].

A detection method for blurred regions was proposed based on local pixel-level metrics, where the detection AUC was above 0.95[78]. A convolutional neural network was trained as a blur detector, which reduced detection error by 12.3% [79]. However, there is currently a lack of out-of-focus correction methods for WSI. A deblurring method for pathological microscope images was proposed, which proved that the blurring of pathological images could be improved by post-processing [80].

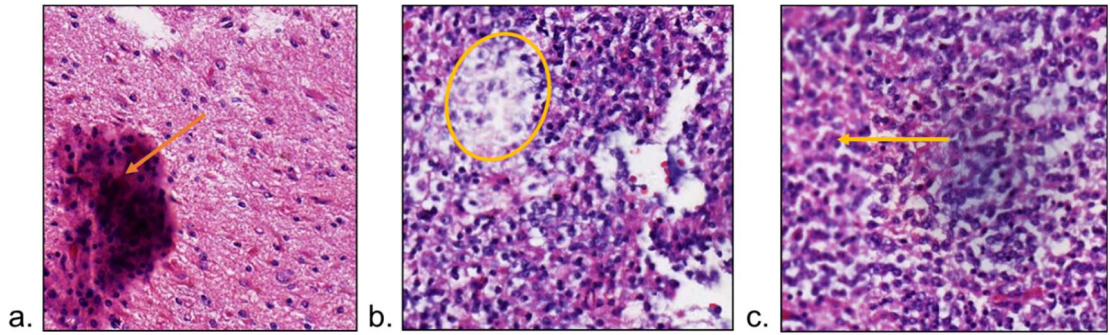

Figure 5. artifacts in histopathology images. (a) the typical tissue folding diagram, where the folded tissue indicated by the yellow arrow is thicker than the surrounding tissue. (b)(c) the regions of out-of-focus, indicated by the yellow circle and the arrow respectively.

### 3.2.3 Data augmentation

PAI based on deep learning often require huge amounts of labeled data to achieve high performance[26]. However, labeled data are often difficult to obtain, especially for some rare diseases. Therefore, the data augmentation techniques are widely used such as cropping, rotating, flipping images[81], changing image contrast and brightness[9].

The generative adversarial network (GAN) to synthesize new data has become promising currently [82]. The generator in GAN generates new synthetic images while the discriminator distinguishes them[83]. The synthetic images increase the number of samples such as rare diseases, which may improve PAI accuracy. 3000 glioma histopathological images produced by GAN were included into the training set for the prediction of the status of the glioma marker isocitrate dehydrogenase, which increased the prediction accuracy from 0.794 to 0.853[84]. The conditional GAN (cGAN) was used to augment training data[13], and the classification accuracy of prostate cancer was improved by 7%, better than 2% of the conventional method.

The synthetic operation reduces the difficulty of collecting a large number of labeled datasets, improving PAI performance. However, histopathological images are rich in structural and texture features [85], causing the GAN to be complex and unstable. The colorectal cancer images generated by cGAN looked blurry and slightly lost the image details[86]. Two pathologists evaluated the quality and authenticity of these images, made conclusion that GAN was effective, and the image boundaries were clear, cytoplasmic colors were correct, but there were still some incorrect features, such as blurred chromatin, lack of nuclear detail and incorrect texture of keratin flakes[87].

Therefore, even if the generated images look visually real, it cannot be included into the dataset before validation[88,89]. The images without further confirmation may adversely affect data distribution and degrade PAI performance.

**3.2.4 WSI review**

In order to ensure the data quality, WSIs are reviewed again before subsequent preparation. Typically, at least two independent review by two senior and experienced pathologists is needed on each WSI. If the review results by two pathologists are agreed with each other, this WSI can be included in dataset, otherwise be discarded [25]. At the same time, the severe color distortions, artifacts, and blurring are checked to make sure they are not present in WSIs. Human review is feasible when making training and testing datasets for PAI. However, in PAI prediction of clinical applications, automatic detection techniques discussed in Section 3.2 should be developed to exclude the WSIs with the presence of severe artifacts, which may cause a significant drop of performance.

**3.2.5 Patch extraction**

Since the image size of 40X WSI can be as high as $100,000 \times 100,000$ pixels, the WSI cannot be input into the graphics processing unit (GPU) for training and testing. Therefore, the small region of interest (ROI) related to the objects such as image region of diseases are usually extracted manually or automatically [25,40].

The height or width of the ROI include up to thousands of pixels, which exceeds the default input of most neural networks. ROI is further segmented into some non-overlapping patches. The patches are usually square areas with dimensions ranging from $32 \times 32$ to $1,000 \times 1,000$ pixels [84]. At a magnification of 20X, the $256 \times 256$[61,90-92] $224 \times 224$[26,93] pixels are common sizes for each patch, matching the input size of most neural networks.

# 4. Annotation

**4.1 Annotation methods**

PAI is mainly based on supervised deep learning, thus requiring a large number of annotated data. Image annotation provides the pairs between images and medical events

(such as diagnostic results) for supervised deep learning. Therefore, a sufficient number of accurately annotated images can improve the PAI accuracy[94,95]. The pathologists use the WSIs for pathological diagnosis, it can be considered that all the WSIs has been annotated. However, the WSIs are so huge, and the target-related regions are usually so small, so the annotations are not accurately related to the regions. Moreover, due to limited computer power, it is usually unrealistic to build PAI directly using the WSI-based annotations[96].

In order to establish an accurate PAI, it is often necessary to further narrow down the annotation, for example the disease location. However, the annotation of WSI is not a simple task, and only professional pathologists can determine the accurate locations. At present, the annotation methods can be classified into fine annotation, weak annotation and sparse annotation.

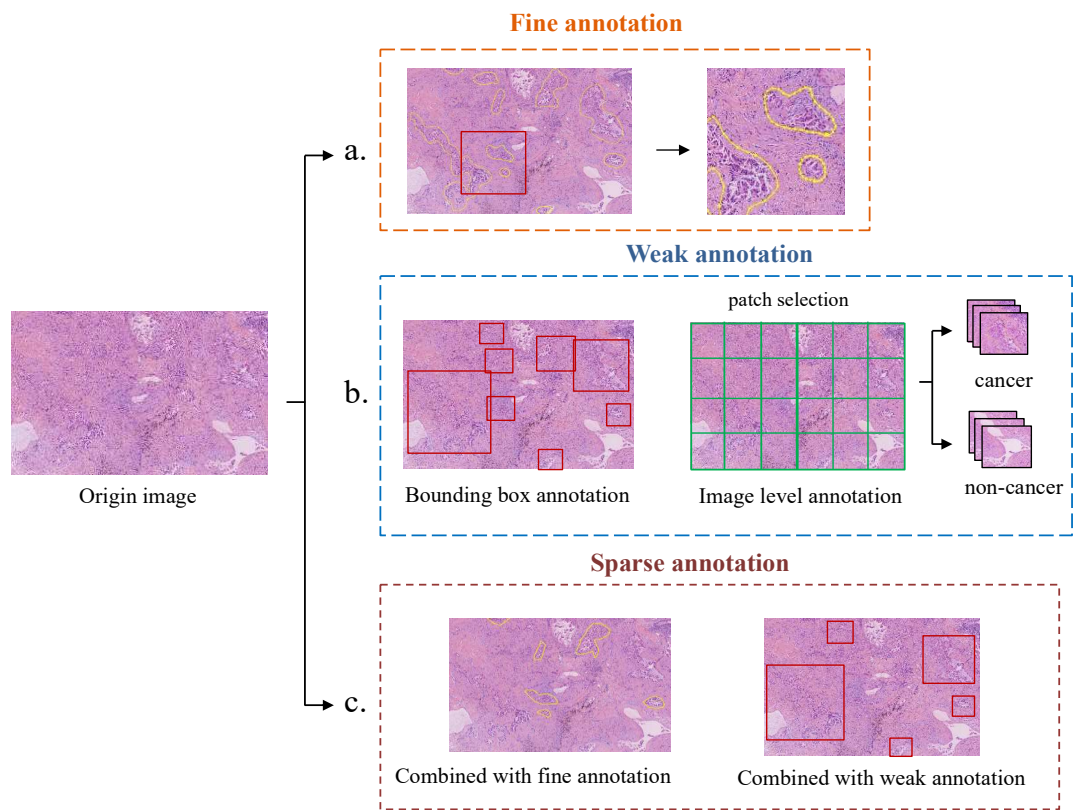

Figure 6. Examples of various annotation methods. (a) Fine annotation, where the boundary between the tumor and normal tissue is depicted carefully. (b) Left: weak annotation using the bounding box, right: weak annotation using a series of non-lapping patches with labels of cancer and non-cancer. (c) Sparse annotation, where only some not all targets in the image are annotated. From left to right, sparse annotation combined with fine annotation, sparse annotation combined with weak annotation.

**4.2 Fine annotation**

The fine annotation (pixel-level annotation) for WSI is commonly used especially for image segmentation[47,97,98]. The location or boundaries of target tissues or cells are precisely depicted, connecting image pixels and targets. Since the kidney tissue was finely annotated pixel-by-pixel, the Dice Coefficient of the glomerular segmentation was 0.95 [99]. The colorectal cancer dataset in the DigestPath 2019 challenge was also finely annotated, the proposed segmentation method achieved Dice Coefficient with 0.7789 and AUC with 1[100].

However, fine annotation is inefficient and time-consuming due to carefully outlining the boundaries/contours or annotating every cell, at the cost of a huge amount of work as well as a lot of time, often requiring many experienced pathologists. Notably, boundaries between tissues are often ambiguous, leading to inconsistencies in labeling among pathologists. Due to the drawbacks mentioned above, the usage of fine annotation is decreasing in addition to the segmentation or measurement for geometric parameters[23,90,101].

**4.3 Weak annotation**

In order to reduce the workload, the annotations of bounding boxes[102], points [101,103] have been used recently, as an alternative to fine annotation. These annotations only point out the target object without providing the exact location or boundaries. PAI by weak annotation can also achieve high performance. Classification accuracy of melanoma images labeled by bounding box was 86.2%, outperforming the dermatologist's accuracy of 79.5%[104]. Since bounding boxes for annotation of dense cell or lesion tissue often overlap each other, point annotation is widely used for cell segmentation tasks. By providing coarse point labels of center-of-mass location of cells, average Dice value of cell segmentation in the ISBI Cell Tracking Challenge dataset 2020 is 0.639[105].

Another common method is to annotate patches. The WSI is firstly divided into a number of non-overlapping patches with the same size manually or automatically. For example, some patches include cancer cells, and other patches only contain normal tissues. The workload of patch-level annotation is often smaller than that of bounding box or point annotation, significantly improving the efficiency of annotation[25,106]. The benign and malignant hepatocellular carcinoma was labeled on the patch, and the obtained Dice of liver cancer cell classification was 0.767, which was slightly higher than the model trained on fine annotation data (the Dice was 0.754)[106]. Moreover, the performance of PAI trained with large-scale weakly annotated data was always much better than that of PAI trained with a small amount of fine annotation data[23].

Recently, many studies based on multi-instance learning have begun to use WSI-level annotation. WSI-level annotation refers to the annotation for the whole image, where the WSI is not divided into small pieces[107,108], so it is more time and effort efficient, and very promising[109]. However, the PAI performance base on WSI-level annotation should be further evaluated. Firstly, most published studies are all done using public not clinical datasets. Secondly, the presented model can often achieve good results only when the disease area in WSI is large. Finally, the presented model may be less effective compared to that of the weak annotation at the patch level.

**4.4 Sparse annotation**

Sparse annotation is the method to reduce the amount of annotation, where only a small number of objects are labeled while other large number of objects not to be labeled. The sparse annotation is often combined with other annotation method to reduce annotation effort. For example, sparse annotation combined with weak annotation was used to label cells by a small number of points, and the accuracy of the trained model was 90.1%, the Dice coefficient was 93.1% [101]. With sparse annotation and fine annotation, the segmentation of gastric tumor images using an iterative learning framework achieved the intersection over union (IOU) of 0.883, and average accuracy of 0.9109 [94].

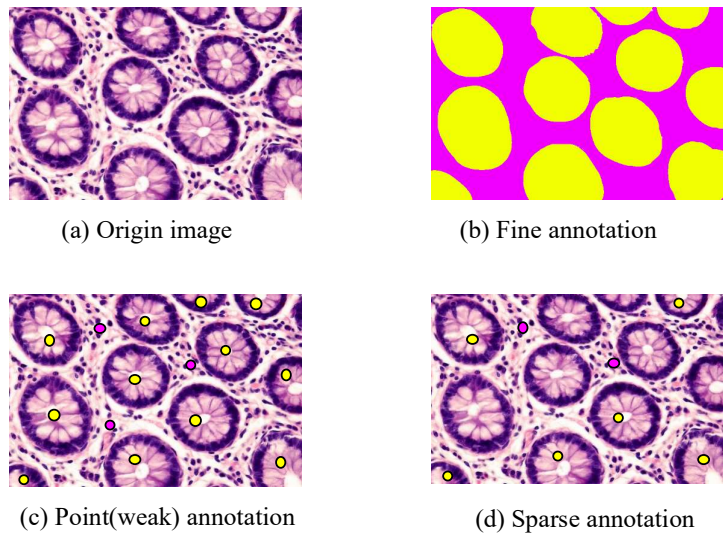

(a) Origin image    (b) Fine annotation

(c) Point(weak) annotation    (d) Sparse annotation

Figure 7 various annotation methods. (a) the original image. (b) fine annotation for the segmentation task, in which the yellow regions represent the cells and the purple represents the background. (c) weak annotation (point labeling), where the yellow dots represent the location of the cells, while the purple dots represent the background location. (d) sparse annotation, where only some cells are annotated by yellow dots.

## 5. Dataset preparation

The collected data is often divided into training set, validation set, and test set for deep learning[59]. The training set is used to generate PAI model, where most of the data is generally used. The validation set is used to select the model i.e., selecting hyperparameter that can achieve the best results while the test set for evaluating the accuracy of the presented model. Figure 8 depicts the role of each dataset in building the model. The raios of the three datasets are not clearly specified, however, in order to obtain a robust PAI, the setting of the training set is crucial.

1) To ensure the independence of each set, those sets should be divided at the patient level. It is guaranteed that the WSIs from the same patient, or patches cut in the same WSI will not appear in different sets [96].

2) The characteristics of the training set greatly affect the performance of PAI, so the training set should cover the subtypes and cell morphology distribution of the disease as much as possible, which is close to the real data distribution of clinical practices.

3) When the dataset is small, most of the WSIs should be used as the training set. While 205 WSIs from a collection of 227 WSIs as the training set, the accuracy of high-grade ovarian is 82%[7]. However, reducing the number of training set from 5,045 WSIs to 1,257 WSIs, the AUC decreased by 5.58% in classification of phosphorylated cell carcinoma[26]. When the collected data is sufficient, the validataion set can be appropriately increased. For example, the 5734 WSIs of lung cancer was collected, of which 3554 WSIs were used for training, 2180 WSIs for evaluation, and AUC of PAI exceeded 0.97[98].

4) In order to evaluate PAI generalization, data from multiple different centers are usually collected. For example, the data from a single center is used for training, while the data form others center is used as an independent test set[14,110]. Some studies incorporated multi-center data into the training set, so that the model was adaptive to the differences of the production and digitization process in these centers. Since the data from 5 centers was used as the training set, the specificity for cervical cancer screening was 93.5%, and the sensitivity was 95.1 % in multi-centers [20].

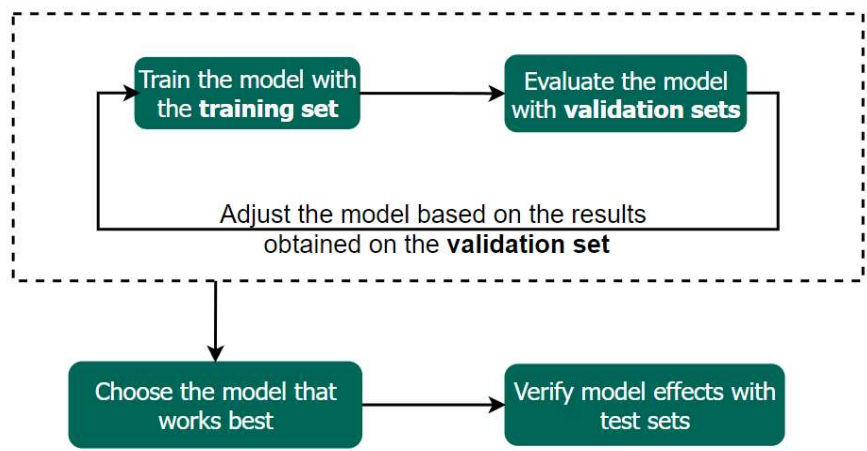

Figure 8. The process of building PAI. First, the training set for training, then validation set for hyperparameters selection by evaluating the performance of the model. After multiple validations, the hyperparameters with the best performance is used for training, and the test set for the evaluation of PAI performance.

## 6. Limitations and improvements of PAI evaluation

PAI evaluation is crucial to demonstrate its clinical effectiveness. The evaluation methods for deep learning fall into internal validation and external validation. When the collected data is insufficient, randomly selecting a portion of the data for testing is commonly used but may cause large fluctuations in model performance. The cross-validation assists in evaluating the performance[98,111], where the dataset is divided randomly into K mutually exclusive subsets. Each time using k-1 subsets for training and the 1 subset for testing, the training and testing are repeated k times respectively. The mean and confidence space of the k cross-validation can eliminate the randomness effect caused by a single division and obtain a more reliable evaluation.

However, the cross-validation is an internal validation due to dataset from the same source. Therefore, the cross-validation can only examine the ability of the model to the samples from the same center, and sometimes the performance may be overestimated[27,112].

The data completely independent of the training set can more accurately evaluates the model[113]. For example, the data from different centers, different scanners, or different production protocols can better test the ability on generalization of PAI. The AUC of the translocation renal cell carcinoma established was 0.886 in internal validation and 0.894 in an independent external dataset, confirming that the presented methods are consistent[65]. A semi-supervised approach for recognition of colorectal cancer achieved the AUC of 0.974 on 12,183 WSIs from 12 medical centers, slightly better than the AUC of 0.969 by pathologists[25].

It is worth noting that the testing set collected by PAI is much smaller than the number of the clinical WSIs, almost impossible to cover all the cellular and histological patterns present in clinic. So, the laboratory evaluation is seriously insufficient to clinical application. After a blinded study on an external testing set, the misdiagnosis of the presented PAI still occurred clinically in 17 sites and 6 cases, including misdiagnosis in detection, grading of prostate cancer and detection of perineural invasion[114]. Additionally, some PAI performance suffered a great decline while using dataset from different countries and medical centers, different patient populations and even pathological scanners. For example, the breast cancer images from different sources scanned by different scanners resulted in a 3% decrease in AUC[77]. Therefore, an overly optimistic view of the actual clinical performance of PAI should be avoided[115]. In order to properly validate the real performance of the model, it is urgent to collect data from different countries [116], and different medical centers to conduct prospective studies[49,112,116,117], thus improving the generalization and robustness of the model.

However, the key to the deployment PAI in clinical practice is the prediction ability to the data unseen in training set[8], but it is very difficult to collect enough multicenter and clinically representative data for evaluation. A feasible method is to evaluate PAI in clinical practices, better comparing the diagnostic performance between PAI and pathologists. Moreover, assessments results such as high diagnostic accuracy do not imply that actual benefits can be provided to patients. Therefore it is a urgent for prospective clinical trial evaluations to demonstrate whether PAI tools can have a positive impact for patients[118].

There are already some guidelines and standards to aid in the assessment of PAI. The Standard Protocol Items: Recommendations for Interventional Trials-Artificial Intelligence (SPIRIT-AI) and Consolidated Standards of Reporting Trials-Artificial Intelligence (CONSORT-AI) are the international standards for clinical trials of AI systems to improve the integrity and transparency of AI clinical trials[118]. PIRIT-AI is an extension of the clinical trial protocol guide SPIRIT 2013 with 15 new entries;And the CONSORT-AI is an extension of the clinical trial reporting guide CONSORT 2010 with 14 new entries[119]. The standards have provided detailed descriptions of AI interventions, instructions、skills、and the integration environment required for use, inputs and outputs, human-computer interaction details, and provision of error case studies. However, SPIRIT-AI and CONSORT-AI mainly focus on supervised learning, and there is still a lack of clarity on how to handle unsupervised and self-supervised learning. Besides, the current standards are primarily image-based and do not yet provide constructive guidance on speech and text types.

## 7. Obstacles and solutions for PAI to clinic

Although data preparation mentioned above can develop a robust PAI, which proves to be effective and accurate in the laboratory setting [59]. However, there are still significant obstacles that make high performance of PAI in laboratory often difficult to reproduce in clinic[120].

Firstly, the number of WSIs in public datasets is very small, containing only typical form of the disease. The images are usually stain-normalized and carefully confirmed by several experienced pathologists[121], but elaborately selected not to cover all the biological and morphological variability. Therefore, other morphologies of diseases that actually exist in the clinic cannot be identified by the PAI potentially. The datasets collected from the clinic can be as many as tens of thousands of WSIs, so the PAI can often achieve better performance. However, these datasets cannot contain enough disease types or subtypes which are rare[122], so the performance of PAI degrades rapidly while these subtypes appear.

Secondly, there are great differences in the production and digitization of slides from multi-centers, as well as color differences in images discussed earlier. The bubbles, tissue folding, and image blurring may reduce image quality and cause a decrease in the performance of PAI. Although normalization techniques such as color correction is helpful, however, when the PAI is used for new center, the normalization method may be retrained.

Thirdly, even if sufficient WSIs can be obtained, the number of annotation is still limited, especially for fine annotation of rare diseases, so the clinical performance of PAI is significantly limited[123]. Weak and sparse annotations can improve the speed of annotation, but they cannot solve the problem of insufficient number of images and lack of representative samples. The GAN can expand the amount of data thus greatly alleviating lack of labeled data. However, GANs have difficulty in correctly generating cases that do not appear in the training set. The unsupervised pre-training, or semi-supervised learning can reduce the amount of annotation required, but their performance may drop significantly if the morphologies of typical disease are not covered[124].

Based on the discussion above, the main obstacle of PAI used in clinic is the dataset such as lack of representative samples that they cannot cover all morphologies. Due to lack of standardization, the differences in production protocols and digitization in other centers have resulted in a decline in the performance of PAI trained on one center. The third obstacle is that the number of annotations is too small. There are three strategies helpfully overcome the obstacles.

### 7.1 Develop digital pathology for clinical diagnosis

Nowadays, most of the samples exist in the form of slides, because they are not digitalized. As discussed earlier, it takes a lot of time for a large number of slides to be digitized for data preparation of PAI. The main reason is that the current diagnosis is based on slides not images. Although high-resolution pathological scanners have begun to be widely used, the number of pixels in WSI is so huge, resulting in extremely high storage and scanning time costs. The clinical application of digital pathology is limited by high cost of slide digitization, so the slide-based diagnosis is widely performed in almost all medical institutions[125].

If digital pathology is used in clinical diagnosis, all the samples are images not slides, thus greatly increasing the number of samples and alleviating lack of representative samples. The key to the development of digital pathology is to reduce the cost of digitization such as storage and scanning time. A low-resolution digitalized method was presented, where the slides were scanned with a 5X lens, 5X images were stored and 40X images were generated in use by super-resolution image technique. While reducing the cost of digitization to one in 64, there was no significant difference in diagnostic accuracy between the high-resolution images generated and the scanned high-resolution images[126]. The method proposed was promising to address the high cost of slide digitization, thereby promoting the development of digital pathology and preparing enough and high quality data for training a clinical-grade PAI.

**7.2 Standardization for slide production and image digitization**

There are many differences in the production protocol and digitization equipment in different centers, thus the images used for training PAI are different from those used in clinical practice. The slide preparation and digitization standardization can reduce the differences, and improve the cross-center performance of PAI. First, the production protocol should be standardized such as the concentration of chemicals used, the quality standards for slides must be clearly specified. Second, the resolution, sharpness or color of pathological scanner should be carefully evaluated based on the WSIs, which ultimately lead to digitization guidelines and industry standards of scanners. Newly generated WSIs before entering PAI should checked by the automated tool on image quality. Therefore, the industry standards of slide production and image digitization are very important to reduce the variance of current clinical data, while the check of image quality should be done before PAI works, helpfully PAI achieving consistent performance in clinical practice.

### 7.3 WSI-level annotation and weakly supervised learning

Due to time cost of pixel-level or patch-level annotations, the amount of labeled data used to train PAI is always insufficient. With the promotion of digital pathology in 7.1, it is not difficult to obtain enough WSI for training. However, there is still major obstacle in labeling a large number of pixels or patches. It is worth noting that WSI-level annotation is naturally present because of pathology diagnoses on WSIs. Weakly supervised learning approach such as multi-instance learning do not require labeled precise locations, so WSI-level annotation may be more suitable.

There are some weakly WSI-based PAIs. The WSIs was coded into a bag, and contextual information of different instances was combined to achieved an AUC of 0.99 and 0.73 in the lung adenocarcinoma classification and lymph node metastasis prediction respectively[109]. A multi-instance learning approach based on clustering and attention was proposed to improve the validity of the data and achieves excellent performance in cancer classification [108]. Recently, the WSI- based PAI for three cancers obtained from the Anatomical Pathology Laboratory Information System (LIS) achieved an AUC greater than 0.98 [23].

Although the performance of PAI trained on WSI-level weak annotation is not yet comparable to the performance of PAI trained on fine annotation, the WSI-level annotation can adapt to the development of digital pathology, where the difficulty of annotation will be greatly alleviated. It is believed that the combination of sufficient annotated data from clinic and WSI-level weakly supervised learning is promising for developing the clinical-grade PAI.

## 8. Conclusion

PAI supports and accelerates the automatic analysis of pathological images, improving the quality and efficiency of pathological diagnosis. Especially in countries lack of pathologists, PAI is a key technology to improve disease diagnosis. This paper reviewed the process of data preparation for PAI carefully, including slide collection, digitization, review, dataset preparation and validation. The PAI is effective and robust, but there are still obvious obstacles for clinical application. This review analyzed the reasons for these obstacles and drew the conclusion that the digital pathology, standardization, and WSI-based weakly supervised learning is promising to build clinical-grade PAI.

# Acknowledgements

This work was benefited by grant supports from the Emergency Management Science and Technology Project of Hunan Province (2020YJ004 and 2021-QYC-10050-26366), and Tongxing Pathology Public Welfare Project from Peking Union Medical College Foundation (G.Y.).